\documentclass[preprintnumbers, floatfix, twocolumn,
preprintnumbers, letterpaper, superscriptaddress,nofootinbib]{revtex4}

\usepackage{graphicx}
\usepackage{microtype}
\usepackage{epstopdf}
\usepackage{amsmath}
\usepackage{amssymb}
\usepackage{subfigure}
\usepackage{hyperref}
\usepackage{url}
\usepackage{xcolor}
\usepackage{color}
\usepackage{mathrsfs}
\usepackage{calrsfs}
\usepackage{amsfonts}
\usepackage{latexsym}
\usepackage{epsfig}

\definecolor{vividviolet}{rgb}{0.62, 0.0, 1.0}
\definecolor{amaranth}{rgb}{0.9, 0.17, 0.31}
\definecolor{palatinateblue}{rgb}{0.15, 0.23, 0.89}
\definecolor{brightpink}{rgb}{1.0, 0.0, 0.5}
\hypersetup{ linktoc=all,
    colorlinks, linkcolor={palatinateblue},
    citecolor={brightpink}, urlcolor={amaranth}
}
\newcommand{\changeurlcolor}[1]{\hypersetup{urlcolor=#1}}
\graphicspath{{Images/}}

\begin{document}

\title{Counting Components in the Lagrange Multiplier Formulation of Teleparallel Theories}

\author{Yen Chin Ong}
\email{ongyenchin@gmail.com}\thanks{corresponding author}
\affiliation{Center for Gravitation and Cosmology, College of Physical Science and Technology, Yangzhou University, Yangzhou 225009, China}
\affiliation{School of Physics and Astronomy,
Shanghai Jiao Tong University, Shanghai 200240, China}

\author{James M. Nester}
\email{nester@phy.ncu.edu.tw}
\affiliation{Department of Physics, National Central University, Chungli 32001, Taiwan, ROC}
\affiliation{Graduate Institute of Astronomy, National Central University, Chungli 32001, Taiwan, ROC}
\affiliation{Leung Center for Cosmology and Particle Astrophysics,National Taiwan University, Taipei 10617, Taiwan, ROC}

\begin{abstract}
We investigate the Lagrange multiplier formulation of teleparallel theories, including $f(T)$ gravity, in which the connection is not set to zero \emph{a priori} and compare it with the pure frame theory. We show explicitly that the two formulations are equivalent, in the sense that the dynamical equations have the same content.
One consequence is that the
manifestly local Lorentz invariant $f(T)$ theory cannot be expected to be free of pathologies, which were previously found to plague $f(T)$ gravity formulated in the usual pure frame approach.
\end{abstract}

\maketitle

\section{Introduction: Gravity As Torsion Instead of Curvature}
\label{intro}

Einstein's theory of general relativity recently celebrated its centenary in 2015, and has so far passed all experimental and observational solar system tests with flying colors. Nevertheless, there remain a few mysteries in astrophysics and cosmology that could be a sign that general relativity might need to be modified on larger scales. On the one hand for galaxies and clusters there are discrepancies that could be explained by large amounts of dark matter or some alternative theory~\cite{1101.1935,1611.02269}.
Another issue is the observation that apparently the expansion of our Universe is currently accelerating~\cite{perlmutter, riess}. A simple explanation is the presence of a positive cosmological constant $\Lambda$, so that the Universe is asymptotically de Sitter in the far future, not asymptotically flat,
but other possibilities cannot yet be excluded.

 These problems motivated the search for a modified theory of gravity, which would agree with general relativity in the regimes
 where
 the latter had been well-tested, but would nevertheless better account for the larger scale observations, maybe giving a more ``natural'' explanation for the cosmic acceleration, \emph{without} the need for $\Lambda$.

The literature has no shortage for various gravity theories, many of which modify general relativity at the level of the action. The most straightforward example is $f(R)$ gravity, in which the scalar curvature $R$ in the Hilbert-Einstein action is replaced by a function $f(R)$. Another class of gravity theories, known as teleparallel gravity, stands out among the rest, for it considers a connection that is curvatureless but torsionful\footnote{We emphasize that torsion, like curvature, is a property of a given connection. Even in a theory with both curvature and torsion, such as the Einstein-Cartan theory, torsion has a clear geometric meaning, and it is best to treat it as such (from the point of view of well-posedness of the evolution equations~\cite{nesterwang}), rather than ``just another field'' coupled to standard general relativity.}. Recall that in general relativity
(GR), the metric compatible Levi-Civita connection has nonzero curvature but vanishing torsion. Gravity is therefore modeled entirely by the effect of spacetime curvature. It may therefore seem rather surprising that there exists a teleparallel equivalent of general relativity (TEGR, or simply GR$_\parallel$), which by construction has zero curvature. For a detailed discussion see~\cite{Pereira.book}. TEGR models gravity as a torsional effect, but is otherwise completely equivalent to general relativity, at least at the action level~\cite{Hayashi, Pereira, Kleinert, Sonester}. Since curvature is identically zero in a teleparallel theory,  there is a \emph{global absolute parallelism}.

 In such theories the torsion tensor includes all the information concerning the gravitational field.\footnote{Here we are considering the standard metric compatible type of teleparallel theory.  There are more general alternatives with both torsion and non-vanishing non-metricity~\cite{Nester:1998mp, Adak:2005cd, Adak:2008gd} which also merit consideration.}  By suitable contractions one can write down the corresponding Lagrangian
density --- assuming an invariance under general coordinate transformations, global Lorentz and parity transformations, and using quadratic terms in the torsion tensor $T{}^\rho{}_{\eta\mu}$~\cite{Hayashi}.
There is a certain combination of considerable interest, the so-called TEGR ``torsion scalar'':
\begin{equation}\label{ts1}
T = \frac{1}{4} T{}^\rho{}_{\eta\mu}T_\rho{}^{\eta\mu} +  \frac{1}{2} T{}^\rho{}_{\mu\eta}T{}^{\eta\mu}{}_\rho - T_{\rho\mu}{}^\rho T{}^{\nu\mu}{}_\nu,
\end{equation}
which is equivalent to the scalar curvature obtained from the standard Levi-Civita connection, up to a total divergence.

A more general quadratic ``torsion scalar'' can be obtained by relaxing the coefficients:
\begin{equation}\label{abc}
T[a,b,c] = a T{}^\rho{}_{\eta\mu}T_\rho{}^{\eta\mu} +  b T{}^\rho{}_{\mu\eta}T{}^{\eta\mu}{}_\rho + c T_{\rho\mu}{}^\rho T{}^{\nu\mu}{}_\nu.
\end{equation}
Only in the case $a=1/4$, $b=1/2$ and $c=-1$ does the theory becomes equivalent to GR.
Finally, $f(T)$ gravity arises as a natural extension of TEGR, if one generalizes the Lagrangian to be a function of $T$~\cite{0812.120, 1005.3039}, see~\cite{1511.07586} for a review. One could in fact have other even more generalized teleparallel theories in addition to $f(T)$ theory.

Here, let us emphasize some aspects of teleparallel gravity theories. As explained above, a \textit{teleparallel} theory is one which is described by a \textit{connection} which is \textit{flat}, i.e., curvature vanishes. On the other hand, one can also formulate a theory described purely in terms of the \emph{frame} (or the co-frame), with no mention of \emph{any} connection. In 4-dimensions this is known as a tetrad theory. It turns out that frame theories and teleparallel theories are essentially equivalent.  People have long understood this to be the case, but we found that there are a couple of subtleties that have not been addressed in the literature. \emph{Firmly establishing this equivalence is the main objective of the present work}.

If one has a teleparallel geometry, starting at any point, one can choose there a basis for the tangent space.  Then one could parallel transport it along any path to every other point in the space.  Since the curvature vanishes, the transport is unique, independent of the path. This constructs a \emph{smooth} global ``preferred'' frame field\footnote{Hence, when trying to solve the equations, one \emph{cannot} hope to get any sensible results by choosing an ansatz frame that is singular, for example the spherical frame (unless one introduces suitably flat connection coefficients, which cancel the singularity in the frame, so that the torsion tensor is smooth.  For a concrete example of this see Section VII in Obkuhov \& Pereira~\cite{OP}).}, in which the connection coefficients vanish, thus one gets a pure frame description --- unique up to an overall constant linear transformation. Conversely, if one has a ``preferred'' smooth frame field, it allows one to introduce a specific parallel transport rule: namely that vectors are transported along paths by keeping their components constant in this frame.  This transport rule is path independent. The associated curvature vanishes. The resulting connection will have vanishing coefficients in this preferred frame.
(This is what is meant by having a connection which is zero.)
Note that geometrically these concepts make sense without any need for a metric tensor (the torsion tensor, as well as the
curvature tensor, can be defined for any connection without using any metric).

Let us suppose we also have a Lorentzian metric (this gives the spacetime a local causal structure),
then there is a distinguished subset of possible teleparallel connections which are metric compatible. With such a connection, if one chooses at one point an orthonormal frame, its parallel transport to all other points will give a global orthonormal frame field. Conversely, given a global orthonormal frame field it determines a metric compatible teleparallel connection. Furthermore any global frame field determines a metric by \textit{defining} the frame to be orthonormal.

One crucial aspect that one has to check for \emph{any} theory of gravity is the number of degrees of freedom it contains. The number is two for general relativity in 4-dimensions\footnote{The number of degrees of freedom for GR in $n$-dimensions is $n(n-3)/2$. In the language of waves, this is the number of polarizations. It is well-known that in 3-dimensions general relativity becomes a topological theory, in which there is no propagating degrees of freedom, and thus also no gravitational waves.}.

Although TEGR has the same degrees of freedom as general relativity\footnote{There are subtleties even in the TEGR case --- in TEGR one physical system is represented by a whole gauge equivalence class: an \emph{infinite set of geometries}, each with its own torsion and distinct teleparallel connection. In the pure frame representation, the gauge freedom representation looks simply like local Lorentz gauge freedom, however it really corresponds to a whole \emph{equivalence class} of teleparallel geometries,
with gauge equivalent torsions.}, a generic teleparallel theory does not.  In the case of $f(T)$ gravity, Miao Li et al.~\cite{Li:2011rn} --- by utilizing the Dirac constraint technique along with Maluf's Hamiltonian formulation~\cite{0002059} ---
concluded that in 4-dimensions there are generically 5  degrees of freedom: namely, in addition to the usual 2 degrees of freedom in the metric, the tetrad would have 3 degrees of freedom. For a more intuitive understanding of why 5 degrees of freedom
could be expected in such a theory, see Sec.2 of~\cite{1412.8383}. {\color{black}(Recently, a Hamiltonian analysis of $f(T)$ gravity was carried out by Ferraro and Guzm\'an \cite{1802.02130}. They claimed that $f(T)$ gravity only contains 3 degrees of freedom, not 5. 
According to our understanding their analysis has some problems, but clarifying  these issues is beyond the scope of the present work.)}

The extra degrees of freedom in $f(T)$ gravity are highly nonlinear, as they do not manifest even at the level of second order perturbation in a FLRW background~\cite{1212.5774}. In fact, it is expected that they will give rise to problems such as superluminal propagation and the ill-posedness of the Cauchy problem in $f(T)$ gravity, i.e., given an initial condition the evolution equations could not uniquely determine the future state of the system. This would be a disaster because it means that physics has lost its predictive power. For comprehensive discussions of this issue, see~\cite{1303.0993, 1309.6461, 1412.8383}. In view of said issue, it is important to further understand the degrees of freedom in $f(T)$ gravity
and other teleparallel theories.

As mentioned, this class of physical theories can be regarded in (at least) two different ways: as a theory formulated purely in terms of an (orthonormal) frame, or as a theory with both a frame and a flat connection.  One could dynamically achieved the flat connection condition by using a Lagrangian multiplier to enforce vanishing curvature.  The frame-connection-multiplier formulation is a particular subclass of the general metric-affine gravity theories, see \S 5.9 in~\cite{MAG}.  As we remarked, it has generally been understood that one can achieve the desired result using this Lagrange multiplier approach.  Upon examination we found that there are some subtle aspects, which have not all been addressed in the existing literature, including~\cite{MAG, 0002022, 0006080, OP}.  We note that the issue is not trivial.  The vanishing curvature constraint depends on the connection coefficient and its first partial derivative.  In Classical Mechanics, it is well known that one cannot in general achieve the desired result by introducing into the action with Lagrange multipliers a constraint which depends on the \emph{time derivatives} of the dynamical variables.  The standard counter-example of such  a \emph{non-holonomic} constraint is ``rolling without slipping'' (for discussions see~\cite{Goldstein} pp 14--16 and~\cite{BatesNester2011}).  Likewise in field theory, one cannot \emph{in general} introduce via Lagrange multipliers constraints which depend on the derivatives of the field, however sometimes this does produce the desired result.  \emph{We do not know of any general results,  so we need to check each case carefully}.

\section{The Lagrange Multiplier Approach}
\label{lagrangemultiplier}

The representation in terms of a \emph{non-vanishing} teleparallel connection may give some insights.
Enforcing vanishing curvature via a Lagrange multiplier has been treated in many sources including Kopczy\'nski~\cite{Kop}, Hehl et al.~\cite{MAG} and Blagojevi\'c~\cite{A}. See also ~\cite{0002022} and ~\cite{HHSS}. This can be done even for the most general metric-affine gravity theory
or for the \emph{a priori} metric compatible case such as $f(T)$ gravity. Our formulation here will essentially be like that of the Obukhov-Pereira metric-affine formulation~\cite{OP}.

It is straightforward to restrict that approach to our needs
by completely eliminating the metric using orthonormal frames.
There are interesting technical details about how the number of independent components of the dynamical equations work out so that this approach is {\em equivalent} to the approach with {\em a priori} vanishing connection. The equivalence has been, until now, not explicitly shown at this level of detail, although most of the underlying ideas were implicit in the earlier foundational works of Blagojevi\'c and Nikoli\'c~\cite{0002022}, as well as those of Blagojevi\'c and Vasili\'c~\cite{0006080}, and Obukhov-Pereira~\cite{OP}. The Lagrange multiplier formulation was also mentioned in a more recent work by  Golovnev, Koivisto, and Sandstad~\cite{1701.06271}, but the counting of the number of components was not carried out. We will demonstrate the equivalence in this section. However, let us first clarify what it means to \emph{not} set the connection to be zero.

In the usual formulation of $f(T)$ gravity, the Weitzenb\"{o}ck connection is defined by
\begin{equation}
\overset{\mathbf{w}}{\Gamma}{}^\lambda{}_{
\nu\mu} = \tilde{e}_A^{~\lambda}\partial_\nu \tilde{e}^A_{~\mu}.
\end{equation}
This expression actually corresponds to a very specific choice of frame in which the frame connection coefficient, often referred to as the \emph{spin connection}, vanishes --- hence we have used $\tilde{e}$ to denote such a preferred
\emph{orthoparallel} frame
(Kopczy\'nski~\cite{Kop} called such frames OT, standing for ``orthonormal teleparallel'').

However, the Weitzenb\"{o}ck connection is well-defined even if we keep the frame connection nonzero~\cite{1510.08432, Pereira.book}:
\begin{equation}
\label{Weitzenb1}
\overset{\mathbf{w}}{\Gamma}{}^\lambda{}_{
\nu\mu}
=e^\lambda{}_A \partial_\mu
e^A{}_\nu+e^\lambda{}_A\omega^A{}_{B\mu} e^B{}_\nu,
\end{equation}
where $\omega^A{}_{B\mu}$ is the frame
connection coefficient defined via $\omega^A{}_B=\omega^A{}_{B\mu} d x^\mu$. In this work, Greek indices $\left\{\mu,\nu,\cdots \right\}$ run over all spacetime local coordinates, while capital Latin indices $\left\{A,B, \cdots \right\}$ refer to the orthonormal frame.
We remark that this formula is not special to the Wetzenb\"ock connection.  It takes any connection components $\omega$ in the frame with upper case Latin indices to the components of the same connection in a frame with Greek indices (which are holonomic here).  There is in general no special restriction on the connection.
For our purpose, $\omega$ corresponds to a flat, Wetzenb\"ock, connection but need not vanish.

One could then calculate the torsion tensor, the torsion scalar $T$, the
action given the explicit form of the function $f(T)$, and the field equations, using the above
Weitzenb\"{o}ck
connection (\ref{Weitzenb1}). For instance the  torsion tensor now reads
\begin{equation}  \label{torsion2}
{T}^\lambda_{\:\mu\nu}=\overset{\mathbf{w}}{\Gamma}{}^\lambda{}_{
\nu\mu}-%
\overset{\mathbf{w}}{\Gamma}{}^\lambda{}_{\mu\nu}.
\end{equation}

However we do not gain anything new, since all this just says that the connection 1-form is non-zero if we go to another basis that is different from the orthoparallel
frame.
In fact, we can work in the Lagrange multiplier approach, and see that the degrees of freedom of the theory remains unchanged.

To be more specific, our claim is this:

\begin{quote}
\emph{The amount of information
 in any teleparallel theory of gravity in which curvature is constrained to vanish via a Lagrange multiplier is the same as that in the formulation in which the connection is set to zero a priori.}
\end{quote}

To see this, let us first consider a general Lagrangian density (i.e., a 4-form in 4-dimensions)  of the form\footnote{For simplicity we do not discuss any matter source fields; they do not play an essential role in the issue we are addressing.}
\begin{equation}
\mathcal{L}(g, \theta,
Dg, T, R, \lambda),
\end{equation}
where $g$ is the metric tensor, %$\omega$ is the connection 1-form,
$Dg$ is the covariant differential of the metric, $T$ is the torsion 2-form, $R$ is the curvature 2-form, and $\lambda$ is a Lagrange multiplier, all of which are written abstractly for convenience.
%impose metric compatibility %$Dg=0$  as an \emph{a priori} constraint, and also
We  ``eliminate'' $g$ as an independent variable via
\begin{equation}
g=\eta_{AB} \theta^A \otimes \theta^B, \qquad \eta_{AB}={\rm diag}(-1,+1,+1,+1),
%\quad 0=Dg_{AB}=dg_{AB}-\omega_{AB}-\omega_{BA}.
\end{equation}
where $\theta^A$ is the orthonormal (co-)frame.
The torsion 2-form and curvature 2-form are related to the orthonormal frame and the connection 1-form by
\begin{equation}
T^A = d \theta^A + \omega^A_{~B} \wedge \theta^B = \frac{1}{2}T^A_{~\mu\nu}dx^\mu \wedge dx^\nu, ~~\text{and}
\end{equation}
\begin{equation}
R^A_{~B} = d\omega^A_{~B} + \omega^A_{~C} \wedge \omega^C_{~B} = \frac{1}{2} R^A_{~B\mu\nu}dx^\mu \wedge dx^\nu.
\end{equation}
We also impose metric compatiblity as an \emph{a priori} constraint:
\begin{equation}0\equiv Dg_{AB}:=dg_{AB}-\omega_{AB}-\omega_{BA}=-2\omega_{(AB)}.\end{equation}
Then $\omega^{AB}$ and $R^{AB}$ are \emph{antisymmetric}: $\omega^{AB}\equiv\omega^{[AB]}$, $R^{AB}\equiv R^{[AB]}$.

Working only with covariant objects, the variation of the Lagrangian density is
\begin{equation}
\delta \mathcal{L} = \delta \theta^A \wedge \frac{\partial \mathcal{L}}{\partial \theta^A} + \delta T^A \wedge \frac{\partial \mathcal{L}}{\partial T^A} + \delta R^A_{~B} \wedge \frac{\partial \mathcal{L}}{\partial R^A_{~B}}+\delta\lambda^A_{~B}\wedge \frac{\partial \mathcal{L}}{\partial \lambda^A_{~B}},
\end{equation}
where
\begin{equation}
\delta T^A = D\delta \theta^A + \delta \omega^A_{~B} \wedge \theta^B, ~~\text{and}
\end{equation}
\begin{equation}
\delta R^A_{~B} = D \delta \omega^A_{~B}.
\end{equation}

Hence
\begin{flalign}
\delta \mathcal{L} =& 
d \left(\delta \theta^A \wedge\frac{\partial \mathcal{L}}{\partial T^A} + \delta \omega^A_{~B} \wedge\frac{\partial \mathcal{L}}{\partial R^A_{~B}}\right) \notag\\ 
&+\delta \theta^A \wedge \epsilon_A
+ \delta \omega^A_{~B} \wedge \epsilon_{A}^{~B}
+\delta\lambda^A_{~B}\wedge \frac{\partial \mathcal{L}}{\partial \lambda^A_{~B}},
 \label{variation}
\end{flalign}
where we introduced symbolic names for the Euler-Lagrange variational expressions:
\begin{equation}\label{EA}
\epsilon_A := \frac{\partial \mathcal{L}}{\partial \theta^A} + D\frac{\partial \mathcal{L}}{\partial T^A}, ~~\text{and}
\end{equation}
\begin{equation}
\epsilon_{AB} := \theta_{[B} \wedge \frac{\partial \mathcal{L}}{\partial T^{A]}} + D \frac{\partial \mathcal{L}}{\partial R^{AB}}.\label{EAB}
\end{equation}
% implies that the curvature 2-form is anti-symmetric:
%$R_{AB}=-R_{AB}$.
%\begin{equation}
%0 = Dg = dg - \Gamma g - \Gamma g.
%\end{equation}
%For orthonormal frame, $dg=0$, and so
%\begin{equation}
%\Gamma_{AB} = \Gamma_{[AB]}.
%\end{equation}
Since $\omega^{AB}$ is antisymmetric $\epsilon_{AB}$ is also: $\epsilon_{AB}\equiv\epsilon_{[AB]}$.

Let us consider a local frame gauge transformation $\delta \theta^A = l^A_{~B} \theta^B$, where $l^A_{~B}$, being an infinitesimal Lorentz transformation, is antisymmetric. We have consequently, $\delta \omega^A_{~B}=-D l^A_{~B}$. Since $\delta \mathcal{L}$ is a scalar under this transformation, we have, from Eq.(\ref{variation}), the following identity:
\begin{flalign}
0 \equiv & ~d\left(l^A_{~B} \theta^B \wedge \frac{\partial \mathcal{L}}{\partial T^A} - Dl^A_{~B} \wedge \frac{\partial \mathcal{L}}{\partial R^A_{~B}} \right) + l^A_{~B} \theta^B \wedge \epsilon_A \notag\\&- Dl^A_{~B} \wedge \epsilon_A^{~B}+(l^A_{~C}\lambda^C_{~B}-l^C_{~B}\lambda^A_{~C})\wedge \frac{\partial \mathcal{L}}{\partial \lambda^A_{~B}}.\label{ident}
\end{flalign}

Since
\begin{equation}
Dl^A_{~B} \wedge \frac{\partial \mathcal{L}}{\partial R^A_{~B}} = -d\left(l^A_{~B} \frac{\partial\mathcal{L}}{\partial R^A_{~B}}\right) + l^A_{~B} D\frac{\partial \mathcal{L}}{\partial R^A_{~B}},
\end{equation}
and $d^2 = 0$, we get from Eq.(\ref{EAB}) and Eq.(\ref{ident})
\begin{flalign}
0 \equiv  &d\left(l^{AB} \epsilon_{AB}\right) + l^{AB} \theta_B \wedge \epsilon_A - Dl^{AB}\wedge \epsilon_{AB} \notag\\
&+l^{AB}\left[\lambda_{BC}\wedge \frac{\partial \mathcal{L}}{\partial \lambda^A_{~C}}
-\lambda^C_{~A}\wedge\frac{\partial \mathcal{L}}{\partial \lambda^{CB}}\right].
\end{flalign}
This yields the Noether differential identity:
\begin{equation}\label{Noether}
D\epsilon_{A B} + \theta_{[B}\wedge \epsilon_{A]} 
-2\lambda_{C[B}\wedge \frac{\partial \mathcal{L}}{\partial \lambda^{A]}_{~C}}
\equiv 0,
\end{equation}
which does not depend on any of the field equations being satisfied.

\section{Counting the Components}
\label{dof}

Now let us consider a special case, the teleparallel Lagrangian:
\begin{equation}\label{La}
\mathcal{L}_{\|}(\theta^A, T^A) + \lambda^A_{~B} \wedge R^B_{~A}.
\end{equation}

The concern is the following: do the field equations obtained from Eq.(\ref{La}) contain the \emph{same} amount of physical information --- no more and no less --- as the equations obtained from the coframe Lagrangian $\mathcal{L}_{\|}(\theta, d\theta)$, or equivalently the frame Lagrangian $\mathcal{L}_{\|}(e,\partial e)$? Note that the variation of the Lagrangian in Eq.(\ref{La}) involves variation with respect to the frame, the connection and the multiplier, whereas the coframe Lagrangian involves only variation with respect to the frame. From the first Lagrangian, the multiplier variation would enforce the vanishing of curvature,
which leads to a preferred frame with a vanishing connection; then the frame variation reduces to that obtained from the pure frame Lagrangian.
So the remaining technical issue is whether the equation obtained by variation with respect to the connection could have any ``physical'' content beyond determining the multiplier. To put it differently: does the connection or the multiplier contain any dynamics?

As mentioned, the variation with respect to $\lambda^A_{~B}$ implies flatness $R^A_{~B}=0$. Then there exists a frame in which $D=d$, in which we no longer have local gauge freedom. However, while it is generally believed that imposing flatness via a Lagrange multiplier is equivalent to imposing flatness \emph{a priori}, it is not obvious how the counting of independent components works out to match so well, especially since in the Lagrange multiplier approach there are gauge degrees of freedom. This is what we shall elaborate on now.

The argument is just as easy, and actually more clear, in $n$-dimensions.
  Then $R^A_{~B}$ is a 2-form while $\lambda^A_{~B}$ is an $(n-2)$-form. It is easy to see that the  Lagrange multiplier has some gauge freedom.  Consider the transformation
\begin{equation}
\lambda^A_{~B} \to \lambda^A_{~B} + D\chi^A_{~B}, \label{multi_gauge}
\end{equation}
where $\chi$ is an $(n-3)$-form.
Under such a transformation the Lagrangian in Eq.(\ref{La}) picks up an additional term
\begin{equation}
D\chi^A_{~B} \wedge R^A_{~B} = d(\chi^A_{~B} \wedge R^B_{~A}) +(-1)^n \chi^A_{~B} \wedge DR^B_{~A}.
\end{equation}
By the Bianchi identity, $DR^B_{~A}=0$. Therefore only a total derivative term is added to the Lagrangian in Eq.(\ref{La}), and thus the equations of motion are invariant under this gauge transformation. In other words, we have a gauge freedom that does not allow us to determine $\lambda^{AB}$ completely, but only up to total differential terms. From Eq.(\ref{EAB}) and Eq.(\ref{La}) we find the explicit form for the expression obtained by variation of the connection one-form:
\begin{equation}\label{eab}
\epsilon_{AB} = \theta_{[B} \wedge \frac{\partial \mathcal{L}}{\partial T^{A]}} - D\lambda_{AB}=0.
\end{equation}
This is the only dynamical equation that contains the Lagrange multiplier, and it indeed
 is invariant under the multiplier gauge transformation~(\ref{multi_gauge})
since, schematically, $D^2 \chi \sim R\wedge \chi = 0$.
Our aim is to show that relation Eq.(\ref{eab}) serves \emph{only} to determine the multiplier (as much as it can be determined), and that it \emph{has no other extra dynamical content} independent of (\ref{EA}).

Let us keep track of the number of independent components. Let $n$ be the spacetime dimension, and $N=\binom{n}{2}=n(n-1)/2$ the dimension of the orthonormal frame gauge group ${\text{SO}}(1, n-1)$.\footnote{Here we are considering the metric compatible case using orthonormal frames.  In other teleparallel theories for the frame gauge group ${\text{GL}}(n)$ one would have $N=n^2$ and for ${\text{SL}}(n)$ $N=n^2-1$.}   The number of independent components of the connection 1-form is $Nn$, that of $R^A_{~B}$ and $\lambda^A_{~B}$ is $Nn(n-1)/2$, and that of $\epsilon^{AB}$ is $Nn$. Finally, the multiplier gauge freedom $D\chi^{A}_{~B}$  has $N(n-1)(n-2)/2$ independent components.\footnote{According to the Hodge-Kodaira-de Rham generalization of the Helmholtz decomposition (see, e.g.,~\cite{AP2, Frankel}), locally a differential form can be decomposed into a sum of terms which are in the kernel and the co-kernel of the differential operator $d$, and can be expressed as the differential and codifferential of certain potentials. For a $k$-form in $n$-dimensions, the sizes of these terms are determined by the binomial coefficients $\binom{n}{k}=\binom{n-1}{k-1}+\binom{n-1}{k}$.}
 Thus the field equations can determine of $\lambda$
\begin{equation}
\frac{Nn(n-1)}{2} - \frac{N(n-1)(n-2)}{2} = N(n-1)
\end{equation}
components.
This is the total number of multipliers minus their inherent gauge freedom. It is effectively the number of components of Eq.(\ref{eab}) that serve the purpose of determining the Lagrange multiplier value. Since we are not actually interested in the values of the multipliers, \emph{this is the content of Eq.(\ref{eab}) that can be neglected}.
There are thus $Nn-N(n-1) = N$ components of Eq.(\ref{eab}) that can contain ``physical information'', since they are not involved in determining the multipliers. However exactly this many components are \emph{automatically} satisfied by virtue of the Noether identity in Eq.(\ref{Noether}), 
the teleparallel condition imposed by the multiplier 
and the frame dynamical equation~(\ref{EA}). Indeed, we observe that in the Noether identity it is not $\epsilon^{AB}$ but $D \epsilon^{AB}$ which actually appears. Due to the differential operator $D$, the identity contains, schematically, $D^2 \lambda \sim R \wedge \lambda \equiv 0$. That is,  $D\epsilon^{AB}$ contains the part of $\epsilon^{AB}$ which is entirely independent of $\lambda^{AB}$. This is an indication that the part of Eq.(\ref{eab}) that is independent of $\lambda^{AB}$ is automatically a consequence of the frame dynamical equation (\ref{EA})
and the identity Eq.(\ref{Noether}) --- \emph{it has no independent information}.

Let us say this in another way.  Here are 4 physically equivalent sets of effective dynamical equations:
\begin{flalign}
\omega^{AB}=0,& \quad E^{AB}=0, \\
R^{AB}=0,& \quad E^{AB}=0, \\
R^{AB}=0,& \quad E^{(AB)}=0, \quad D\epsilon^{AB}=0, \\
R^{AB}=0,& \quad E^{(AB)}=0, \quad \epsilon^{AB}=0,
\end{flalign}
where the frame dynamical equation has been written as a 4-form
\begin{equation}E^{AB}:=\theta^A\wedge\epsilon^B.\end{equation}
{Here $E^{(AB)}$ denotes its symmetric part and $E^{[AB]}$ its antisymmetric part.}
Effectively, $E^{[AB]},~D\epsilon^{AB}$ and $\epsilon^{AB}$ (\ref{eab}) contain equivalent physical information.
The key is the Noether differential identity, Eq.(\ref{Noether}), which guarantees that
\begin{equation}
E^{[AB]}=0 \ \Longleftrightarrow \ D\epsilon^{AB}=0.
\end{equation}
In the frame with vanishing connection the second of these equations says that $\epsilon^{AB}$ is closed, then (at least locally)
it is exact---which thus means one can find a multiplier in (\ref{eab}) that makes $\epsilon^{AB}$ vanish.

Thus the Lagrange multiplier approach yields the same number of independent components as the usual approach in which the curvature-free condition is imposed \emph{a priori}.

There still remains a \emph{slight} possibility that the first term on the right hand side of Eq.(\ref{eab}) might, in $n$-dimensions, contain a closed but not exact $(n-1)$-form. Then it might include an extra global condition for the connection-multiplier representation that is not required in the coframe version. \emph{To us this seems unlikely, but we have not yet been able to rule it out for spaces that have a non-vanishing $(n-1)$-cohomology}\footnote{Future works considering explicit examples of 4-dimensional spacetimes with nontrivial 3-cohomology might shed some light on this issue. We propose to study class A Bianchi models (types I, II, VIII, IX), which can all be compactified. In particular, Bianchi type I model can have a 3-torus topology, and type IX can have an $S^3$ topology.  Both of these spacetimes have spatial volume 3-forms that are closed but not exact.}.

Thus \emph{generically} a teleparallel theory has effectively $n^2$ physical dynamical equations $0=E^{AB}=E^{(AB)}+E^{[AB]}$.
Only for the special case of the teleparallel equivalent of GR
the anti-symmetric part vanishes identically:
$E^{[AB]}\equiv0$, leaving $n(n+1)/2$ dynamical equations.

It is important to emphasize at this point that, for TEGR, in the connection-multiplier representation there are \emph{two} local Lorentz symmetries:
\begin{itemize}
\item[(1)] Transforming the frame along with the standard induced connection transformation leaves the action invariant.
\item[(2)] Transforming the frame while keeping the connection fixed changes the action by a total differential.
\end{itemize}
Transformation (1) applies to all teleparallel theories, whereas (2) is obviously is no longer true in the case of a general teleparallel theory, such as $f(T)$ gravity.

\section{Conclusion}
One major advantage of the Lagrange multiplier formulation is that it permits us to use \emph{any} orthonormal frame that corresponds to a metric, since it manifestly preserves local Lorentz invariance. This avoids the important and practical problem of identifying the correct frame compatible with the zero-connection in the usual approach.

Although it has long been argued that this approach is equivalent to the usual frame
 approach which sets the connection to zero \emph{a priori}, we found that there are some subtleties in the counting of the number of components in the Lagrange multiplier approach, which until now have not been fully discussed in detail. In this work we showed that indeed the number of physically significant components for the equations in the Lagrange multiplier formulation agrees with that obtained using the frame approach.

Consequently,
 a manifestly local Lorentz invariant $f(T)$ theory cannot be expected to be free of the pathologies, which were previously found to plague $f(T)$ gravity formulated in the usual pure frame approach. Nevertheless, the Lagrange multiplier teleparallel formulation might shed some light on the properties of the extra degrees of freedom and the ``remnant symmetry'' discovered in~\cite{1412.3424} (which was further discussed in~\cite{1412.8383}).

\section*{Acknowledgement}
YCO acknowledges the support from National Natural Science Foundation of China (No.11705162), and Natural Science Foundation of Jiangsu Province (No.BK20170479).
The authors thank Manos Saridakis, Martin Kr\v{s}\v{s}\'ak, and Huan-Hsin Tseng for related discussions.
YCO also thanks Martin Kr\v{s}\v{s}\'ak for his hospitality during the ``Geometric Foundations of Gravity'' conference in Tartu, Estonia, during which this work was finalized.
He acknowledges the China Postdoctoral Science Foundation (grant No.17Z102060070), which supported this travel.

%%%%%%%%%%%%%%%%%%%%%%%%%%%%%%%%%%%%%%%%%%%%%%%%%

\end{document}